# Discovering a Regularity: the Case of An 800-year Law of Advances in Small-Arms Technologies

Alexander Kott, U.S. Army Combat Capabilities Development Command Army Research Laboratory, Adelphi, MD, USA

Philip Perconti, U.S. Army Combat Capabilities Development Command Army Research Laboratory, Adelphi, MD, USA

Nandi Leslie, U.S. Army Combat Capabilities Development Command Army Research Laboratory, Adelphi, MD, USA

Corresponding author: Alexander Kott, CCDC Army Research Laboratory, 2800 Powder Mill Road, RDRL-D, Adelphi, MD 20783-1197; email: alexkott@yahoo.com

Declaration of Interests: none.

Funding: none.

Disclaimer: The views presented in this paper are those of the authors and not of their employers.

## Abstract

Considering a broad family of technologies where a measure of performance (MoP) is difficult or impossible to formulate, we seek an alternative measure that exhibits a regular pattern of evolution over time, similar to how a MoP may follow a Moore's law. In an empirical case study, we explore an approach to identifying such a composite measure called a Figure of Regularity (FoR). We use the proposed approach to identify a novel FoR for diverse classes of small arms -- bows, crossbows, harquebuses, muskets, rifles, repeaters, and assault rifles -- and show that this FoR agrees well with the empirical data. We identify a previously unreported regular trend in the FoR of an exceptionally long

duration – from approximately 1200 CE to the present –and discuss how research managers can analyze long-term trends in conjunction with a portfolio of research directions.

## 1. Introduction and Motivation

Trend-based approaches are important tools for technology analysis. For example, exponential trends in technology growth have been observed for many decades and have been explained in part by action-reaction behaviors of market competitors [Seamans 1969]. Such trends are found in multiple technologies; e.g., Nagy et al. [2011] describes exponential laws for 62 different technologies. A particularly well-known example is Moore's law [Moore 1965, Schaller 1997, Gelsinger 2006, Koomey et al. 2011], so much so that exponential trends in technology are often called simply a generalized Moore's law. Although exponential models tend to be popular, superexponential models (e.g., Nagy et al. [2011] and Sandberg [2010]) as well as other alternatives for study of technology trends (e.g., Sood et al. [2012]) have also been studied.

To identify a technology trend of a class of devices, one needs to formulate a measure, a quantity that describes the state of technology at a given time, and exhibits a trend over time [Martino 1993a, p.93; Coccia 2005]. Commonly, the literature on such trends assumes a measure of performance (MoP) and plots the measure as a function of time, e.g., the year when the technology first appeared. Often, such plots exhibit regularity in the sense that all MoP points fall close to a single curve depicting the growth of the MoP over time, despite representing different products from different manufacturers and different design approaches. In addition, the curve often offers intuitive appeal in terms of the explainability and parsimony of the underlying equation, e.g., an exponential or superexponential.

Our research, however, focuses on those technologies where formulation of a MoP does not appear amenable to either an intuitive assumption or an approximate analytical model. Consider for example the long history of a complex group of several technology families with a common function – infantry small arms. In this paper, we focus on those small arms that meet the following criteria: (a) deliver projectiles at an adversary; (b) used by a foot soldier; (c) serve as a primary and not a supporting weapon, such as pistols and submachine guns; and (d) do not require more than a single individual to operate, e.g., machine guns that require an assistant to carry the ammunition. Because the timeframe we use for identifying the technology trend is over 800 years long, such weapons include bows, crossbows, harquebuses, muskets, rifles, repeating rifles, and assault rifles.

Although such technologies are the subject of in-depth engineering disciplines and literature, e.g., McCoy [2012] and Carlucci and Jacobson [2018], we are not aware of an established MoP of a kind that would be suitable for technology trend analysis. The great challenges in defining meaningful measures of performance or effectiveness for weapon systems in general and infantry small arms in particular are discussed in multiple studies, e.g., Brown [1995], Stockfisch [1975], and Crenshaw [1986]. It is difficult to see a property of small arms, or a combination of properties, that exhibits a consistent growth typical of technologies that comply with a Moore's law or its variations.

For example, as seen in Fig.2, the muzzle kinetic energy of small arms (the kinetic energy of a projectile as it departs the weapon [Kott 2019]) changed over time in a complex manner that is hardly consistent with an exponential, superexponential, or other obvious regular trend. Other characteristics of small arms, such as the rate of fire or the effective range, exhibit different temporal dynamics but also without an obvious regularity.

In such cases, therefore, when a MoP is not determined, it would be desirable to find some other measure – let us call it a Figure of Regularity (FoR) – that is a function of a technology's attributes and exhibits a regularity often found in temporal evolution of an MoP. We would like to find a FoR that follows – like a MoP – a temporal pattern plausibly expected for the evolution of technology, and that is useful for technology analysis and forecasting. We envision a FoR that is a mathematical expression that combines a number of a technology's attributes that are considered significant by domain experts. To that end, we seek an approach that would allow us to derive or model such a FoR from the historical data associated with the technology.

Our notion of a FoR revisits the pioneering idea of Alexander and Nelson [1973], who found a useful correlation between the date of introduction of a product (in their case, jet engines) and a weighted sum of logarithms of the product's attributes. That sum served in effect as a substitute for a performance measure, as Martino [1993b] suggested. In this paper, we refer to this idea as Alexander and Nelson Regression (ANR). We generalize ANR, as discussed later in the "Problem Definition and Solution Outline" section.

One may wonder whether a FoR represents a MoP. After all, one might argue, it does behave like a MoP in exhibiting a regular growth over some period of the technology's history. This is a complicated issue, particularly because the very term "measure of performance" lacks a rigorous, formal definition. We elect to keep this question for future research.

Thus, our broad research question is: what analysis methods can uncover long-term regularities in technologies that lack an accepted MoP? A narrower question, the focus of this paper, is whether a long-term regularity can be discerned using a generalization of the ANR method? We explore this question empirically, in a particularly challenging context of technologies spanning 800 years that are functionally related but as technologically diverse as medieval bows and modern assault rifles.

The contributions of this paper are four-fold.

First, we formulate and motivate the problem of deriving – from empirical, historical data – the form in which a technology's FoR depends on key attributes of the technological artifact, together with determining the model of growth of FoR over time. To our knowledge, such a formulation of the problem has not appeared in literature.

Second, we propose an approach to solving this problem and empirically explore its efficacy in a complex case of small arms, ranging over diverse families and epochs of technologies.

Third, in this case study, we identify a previously unreported (to our knowledge) superexponential trend in a small arm's FoR over a remarkably long duration. This implies that other cases may exist where consistent trends extend over periods that far exceed the periods mentioned in prior literature.

Fourth, we show the application of the resulting solution to common challenges of strategic technology management, e.g., of forecasting the long-term (e.g., 30 years) increase in FoR and assessing whether certain research directions might support such an increase.

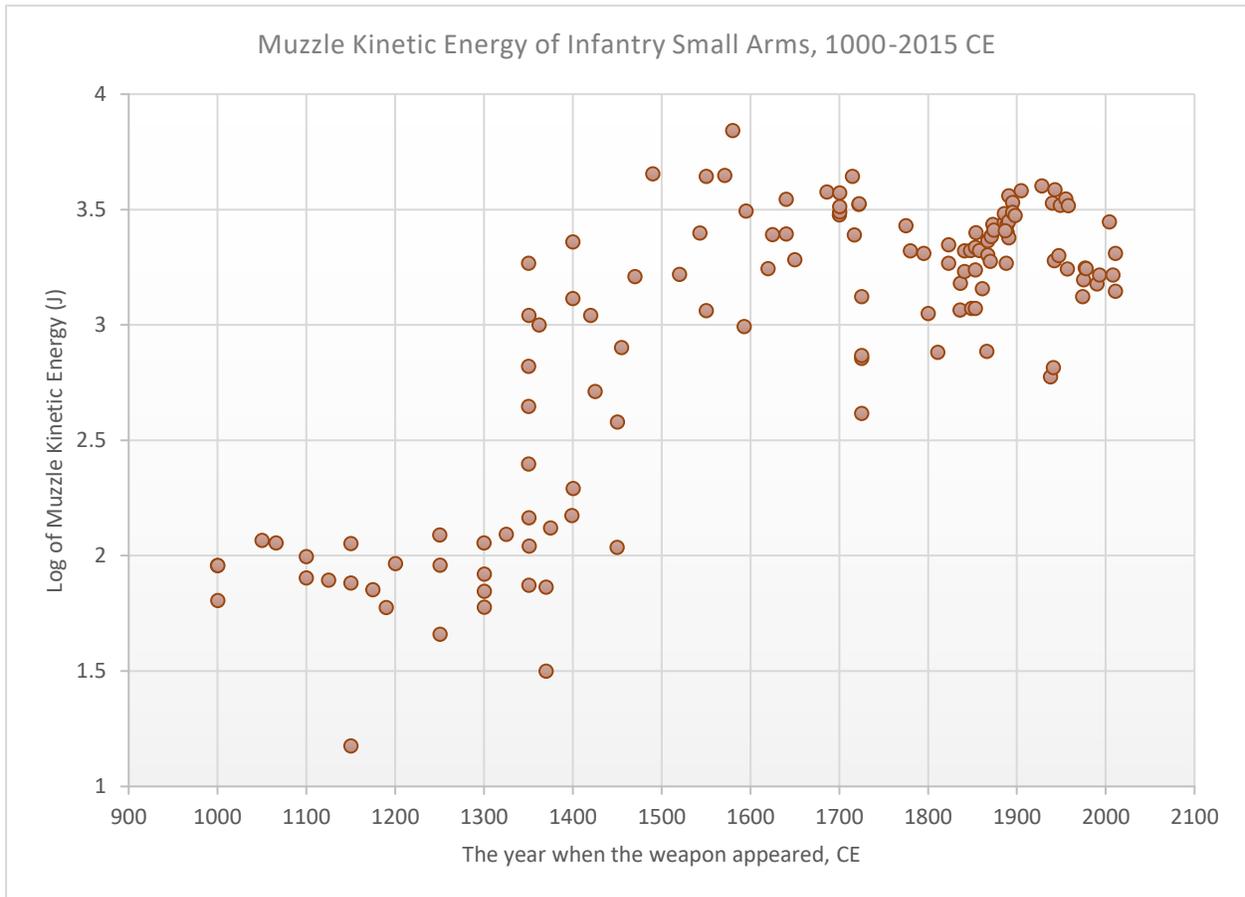

*Figure 1. Over the last 10 centuries, the kinetic energy of small-arms projectiles (arrows, bolts, balls, bullets) exhibited a complex pattern of changes. Although an important characteristic of small arms, it does not seem to follow any exponential or superexponential trend.*

## 2. Problem Definition and Solution Outline

We now give a problem formulation and a solution approach to using empirically observed data to discover a FoR of a given technology domain simultaneously with the technology's temporal evolution. Our approach generalizes the ANR method by allowing additional temporal growth models, other than strictly exponential in time as in the original ANR. Alexander and Nelson [1973] were the first to note and study a correlation between the date of a technology introduction, e.g., the year of appearance of a certain jet fighter or an engine, and a function that combines several attributes of the technology. Their approach is equivalent to use of model A in the "Models of Temporal Dynamics" section of this paper, corresponding to the hypothesis of a linear regression in the time-series analysis. Like them, we find it convenient to use the general form of a FoR, as described in the "Modeling the FoR" section. Unlike them, however, we formulate a more general problem (Eq. 1) with the explicit recognition that different laws of temporal dynamics may need to be considered, and the search for the most fitting the temporal dynamics should be conducted simultaneously with the search for the FoR model. Different models of temporal dynamics may correspond to different models of a FoR's dependency on the technology's attributes or different parametrizations of an FoR model. An a priori assumption of an exponential temporal dynamics model could lead to poor results. In fact, the implicit assumption of original ANR – that the logarithm of the FoR is a linear function of time – yields a misleading form of a FoR, as we demonstrate later.

The basic mathematical problem considers a real-valued vector $x$ for the attributes of technology artifacts measured or observed at a set of times $t_i$, with time-series data $\{x(t_i), i = (1, ..., M)\}$, for assessing attributes, where $x_i = x(t_i)$. In our case study, these variables characterize small arms are projectile mass, maximum rate of fire, maximum effective range, and muzzle velocity, described further in later sections. In addition, we define a metric $f$ (also referred to as the FoR throughout our paper) to provide a single composite measure of small-arms features modeled such that $f: \mathbb{R}^n \to \mathbb{R}$, described as

$$f(x) = k \prod_{i=1}^{n} x_i^{\alpha_i}, \qquad (1)$$

where $k > 0$ and $\alpha_i \geq 0$ for $i = 1, ..., n$ are constants to be explored further in Section 4.2.

We also seek to predict how $f$ changes with time. Examples of the observations of small-arms technology data appear in Table 1. We define a set of models (in our case study, a set of five) $F = \{\hat{f}^k(t_i) ; i = (1, ..., M), k = (1, ...,5)\}$, where $\hat{f}_i^k = \hat{f}^k(t_i)$ approximates $f$ at each time step $t_i$, and $\hat{f}_i^k$ provide alternative models of the FoR over time.

One possible hypothesis used in our case study is that the FoR increases exponentially with time $t_i$, which is the original ANR model, and can be approximated with the exponential model, $y_i^1 = \hat{f}^1(t_i) = \exp(\theta_2 + \theta_1 t_i)$, where $\theta = (\theta_1, \theta_2) \in \mathbb{R}^2$ are the fixed parameters of the model, $\ln(\hat{f}^1(t_i))$ is a linear regression model, and the maximum likelihood estimate of $\theta$ is given by $\hat{\theta}$. In Section 4.2, we define four additional models as alternatives to this exponential trend.

For model selection, we score the alternatives with one of the many statistics based on the maximized log likelihood of the data, $L = \sum_{i=1}^{n} \ln(\Pr_{\hat{\theta}}(y_i^k))$, the Bayesian Information Criterion (BIC), which has a penalty applied to avoid overfitting

$$BIC = -2\ln(L) + 2k\ln(n), \qquad (2)$$

where $k$ is the number of free model parameters and $n$ is the number of data points [Friedman et al., 2001]. Under the Gaussian model, BIC for a model is estimated by

$$BIC = \frac{n}{\hat{\sigma}_e^2}\overline{err} + \frac{k}{n}\hat{\sigma}_e^2 \ln(n), \tag{3}$$

where $\hat{\sigma}_e^2$ is the variance and the error $\overline{err}$ is defined as $\overline{err} = \frac{1}{n-1}\sum_{i=1}^{n}(y_i^k - \hat{f}_i^k)^2$ [Friedman et al., 2001]. In addition, we explore other goodness-of-fit statistics, mean absolute percentage error (MAPE) and $R^2$.

In the "Methods and Materials" section, we solve such a problem for a specific case of small-arms technology. Let us use an example from that section in order to outline and illustrate an approach for solving the above problem.

Step 1. Use domain-specific expertise to identify the most significant attributes $\{x_i, i = (1, ..., M)\}$ for the technology in question using time-series data. In the "Methods and Materials" section, we describe four attributes selected as sufficiently characterizing the important and most broadly applicable properties of small arms. Table 2 shows examples of the data points.

Step 2. Using prior work and expert knowledge, hypothesize a set of models $F$. For example, in the "Methods and Materials" section, we hypothesize five models that describe how a FoR of small arms might have varied over the period of 1200–2015 CE, each model with a set of parameters.

Step 3. Similarly, hypothesize how $f(x)$ depends on attributes. For example, in the "Methods and Materials" section, we hypothesize a model that describes how the FoR of small arms depends on the four attributes determined in Step 1.

Step 4. Use the least-squares approach (or minimize another appropriate loss function) to find the model parameters $\hat{\theta}_p^k$ for the k[th] model with p parameters. In the "Fitting the Models" section, for example, we do so for the set $F$ formulated in Step 2.

Step 5. Compute the goodness-of fit statistics to select the best-fit model in $F$. Assess the overall goodness-of-fit statistics for the alternative models. For example, in the "Discussion of Results" section, we assess the BIC, and the lower BIC value is assumed to be one of the criteria for selecting the best-fit model.

## 3. Prior Work

There is extensive research on exponential and superexponential models for technology trends, their empirical evidence, and theoretical explanations. Seamans [1969] proposes a mechanism – a sequence of competitive actions and reactions – that lead to an exponential trend. Martino [1993a] provides a comprehensive monograph covering many topics of technological forecasting including exponential trends, and Martino [1993b] presents and compares approaches to forming composite measures of technologies. Nagy et al. [2013] demonstrates the broad applicability of exponential trends to multiple technologies. Superexponential models, e.g., Nagy et al. [2011] and Sandberg [2010], may also apply. These works generally rely on an *a priori* assumption of a certain MoP, which may not be available in some technology domains. Our approach, instead, uses empirical data to derive a FoR that behaves in a regular fashion like a MoP might.

The ANR approach is the pioneering idea of Alexander and Nelson [1973]. It has been elaborated upon by Martino [1993a, 1993b] and more recently discussed in Inman et al. [2006]. Our work differs by generalizing ANR to temporal growth models other than exponential.

The near-millennium scope of our case study fosters connections within broad histories of military technologies such as Dupuy [1982], Van Creveld [2010], and Knox and Murray [2001]. Even more relevant – because they often provide more quantitative details – are historic studies of technology trends focused on a particular period, for example, Gabriel and Metz [1991], Williams [2003], Duffy [1988], Lewis [1956], Kerr [2015], and Lawrence [2017]. Of special value are studies focused on the characteristics of particular classes of weapons, such as McLachlan [2010], Phillips [1999], [Reid 2016], Krenn et al. [1995], and Roberts et al. [2009]. None of these works, however, identify a long-term quantitative regularity in characteristics of small arms, as this paper does.

Because this paper is somewhat unusual in its aim to provide a strategic, long-term analysis of technology, we should mention examples of long-range forecasts such as Albright [2002] or forecasts specifically about military technologies, such as Newman [1996], Vickers [1996], and O'Hanlon [2000]. It has been shown that the accuracy of such forecasts can be quite high, about 70–80% [Kott and Perconti 2018]. In focusing on the year 2050 as our forecast horizon, our paper shares its target with other works that range from broad envisioning of warfare technology as in Singer [2009], to the focused projections of selected scientific breakthrough in Kott et al. [2018].

## 4. Methods and Materials: the Case of Small-Arms Technologies

### 4.1 The Data

The data are extracted from Kott [2019]. All data meeting the criteria stated in section 1 are considered in this study, with no exclusions. The dataset comprises over 120 data points, where each data point represents a distinct weapon (e.g., a Brown Bess musket; see Table 1). The data are primarily for Western European and U.S. weapons, as this is the scope of our research. The entire collection of the data, the sources, and the approaches to interpreting or approximating the data are available and discussed in detail in Kott [2019]. Table 1 gives examples of a few data points in that collection.

*Table 1. Examples of data. Full set of data and notes on sources are in Kott [2019].*

| Weapon | Year (CE) | Projectile mass (kg) | Max rate of fire (1/min) | Max effective range (m) | Muzzle velocity (m/s) |
|---|---|---|---|---|---|
| Longbow | 1180 | 0.1023 | 5 | 75 | 47 |
| Crossbow 13c | 1250 | 0.0840 | 2 | 75 | 45 |
| Handgonne | 1350 | 0.0380 | 0.5 | 25 | 149 |

| Harquebus | 1455 | 0.0278 | 1 | 50 | 240 |
| Wheel lock musket | 1595 | 0.0300 | 1 | 75 | 456 |
| Brown Bess musket | 1722 | 0.0321 | 3 | 75 | 457 |
| Berdan rifle | 1870 | 0.0198 | 7 | 270 | 437 |
| M27 assault rifle | 2008 | 0.0041 | 700 | 550 | 900 |

Each data point can be described by $(t_1, x_1), \ldots, (t_M, x_M)$, where each $x_i = (x_{i1}, x_{i2}, x_{i3}, x_{i4})$ and $t_i$

is an unevenly spaced discrete time series representing the approximate year in which the weapon was introduced or designed. We limit the period under consideration to 1000–2018 CE. In most cases, sources exist that report the date of the weapon's design or introduction into service, but in some cases, we had to resort to assumptions.

In the data, $x_{i1}$ is the muzzle velocity or the projectile velocity at the moment of separation from the weapon (e.g., the arrow velocity as it exits the bow or the bullet velocity when it exits the muzzle). We denote the maximum effective range with $x_{i2}$, i.e., the distance at which an infantryman can fire the weapon with an acceptable probability of hitting and disabling the targeted adversary. The mass of the projectile $x_{i3}$ is another important feature in the data. Finally, the $x_{i4}$ represents the maximum rate of fire as a feature in the data, i.e., the maximum number of projectiles per minute that an infantryman can fire from the weapon. Detailed discussion of domain-specific significance of these attributes is found in Kott [2019].

## 4.2 Modeling the FoR

We consider the four attributes introduced previously as important for modeling the FoR of infantry small arms. We also see them as meaningful regardless of the technological features used to achieve those functional characteristics, i.e., applicable across families of technologies as diverse as longbows and modern assault rifles. To put this in the terms of Eq. 1, this is the vector of attributes $x_i$.

Now we need to formulate a set of models $F$, where each model $f_i$ describes how the technology's FoR depends on the attributes $x_i$. To constrain the scope of this research, we consider the single model in Eq. 1. The form of the model is similar to the one in Alexander and Nelson [1973] and Martino [1993b].

## 4.3 Models of Temporal Dynamics

Our next step is to hypothesize the set of models $F$ that describe how the technology's FoR changes with time. Each model $\hat{f}_i^k$ is parametrized with a set of parameters $\hat{\theta}_j^k$, where $i = 1, \ldots, m$, $j = 1, \ldots, p$, and $k = 1, \ldots, 5$. To hypothesize the following five models, we follow primarily Nagy et al. [2013],

### Model A – Exponential Growth

This model $\hat{f}_i^1$ is the most common form of the Moore's law [Nagy et al. 2013]. It is the linear regression model that also underlies the original ANR. This form of temporal dynamics implies the hypothesis that the FoR increases by a constant fraction per unit of time such that $\hat{f}_i^1 = \exp(\theta_2 + \theta_1 t_i)$.

### Model B – "Quadratic Exponential"

If the rate at which the FoR increases is a fraction that is not constant, but increases over the time, then a parsimonious hypothesis would be that the fraction is a linear function of time. This leads us to suggest the model $\hat{f}_i^2 = \exp[\theta_2 \cdot (t_i - \theta_1)^2]$.

### Model C – "Cubic Exponential"

Continuing the same logic, if the fraction is a quadratic function of time, then one other model could be

$$\hat{f}_i^3 = \exp[\theta_2 \cdot (t_i - \theta_1)^3].$$

### Model D – Double Exponential

Following the hypothesis of Kurzweil [2001], we may assume that the fraction itself grows exponentially over time. Then, a model could be $\hat{f}_i^4 = \exp[\theta_1 + \exp(\theta_2 + \theta_3 t_i)]$.

### Model E – Piecewise Exponential

This model is inspired by in part by the piecewise model of Nagy et al. [2011] and in part by Lienhard [1979] who argued that a drastic change in the rate of technological dynamics occurred in the year 1832. The year 1832 appears plausible for the purposes of our study, because the period between the 1820s and 1840s witnessed such revolutionary developments in small-arms technology as the percussion cap, the Minié ball, and the Dreyse gun. The model is $\hat{f}_i^5 = \begin{cases} \exp(\theta_1 + \theta_2 t_i), \text{ if } t_i \leq 1832 \\ \exp(\theta_1 + \theta_2 t_i), \text{ if } t_i > 1832 \end{cases}$, where $\theta_4 > \theta_1$.

## 4.4 Fitting the Models

For each of the five models of temporal dynamics, we performed linear regression on the data, seeking the best least-squares fit between the values of the time-series model and the FoR model. This yields the values of parameters $\theta_k$ for the FoR temporal dynamics model. In order to give the FoR function a domain-relevant interpretation, we aim to relate the FoR to the kinetic energy of the projectile and therefore scale all parameters so that $\alpha_1$ = 2.0 (more on the interpretation later). Furthermore, we scaled all parameters so that $\hat{f}_i^k$ = 1.0 at $t_i = 1200$. This scaling does not affect the goodness of fit.

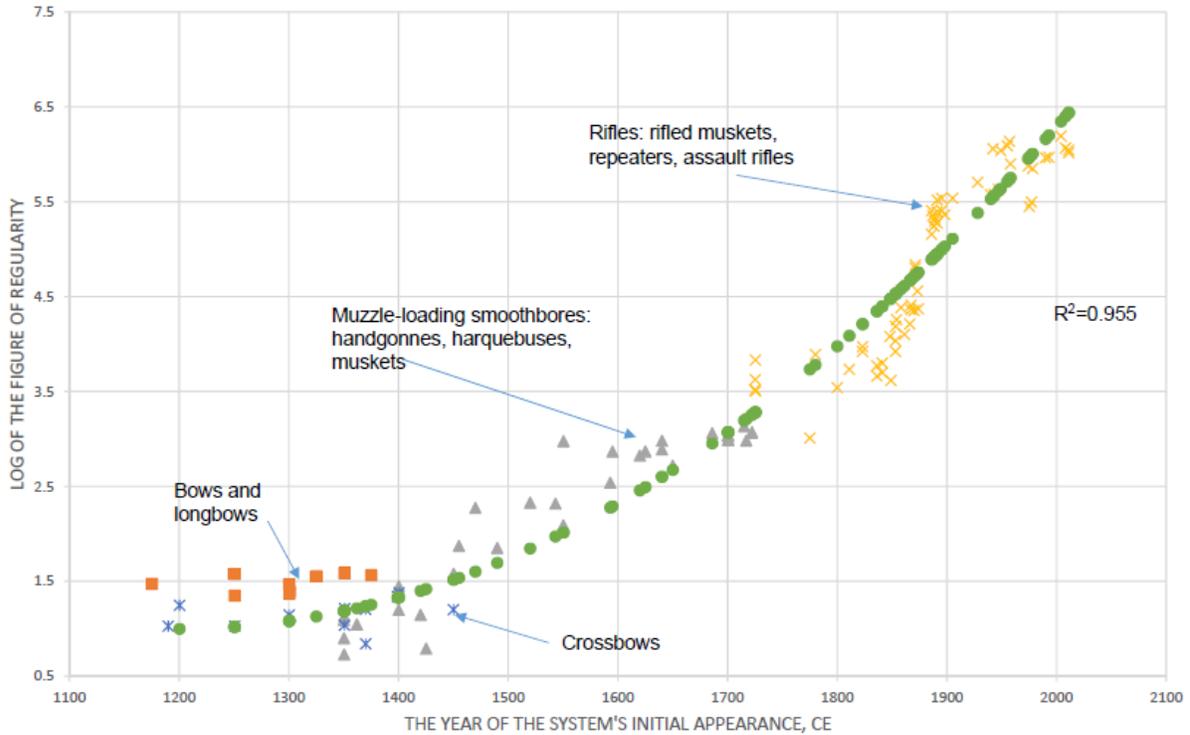

Figure 2. The logarithm of the composite measure, log(FoR), is closely approximated by a quadratic function of time $1.0 + 8.27(10)^{-6} (Year-1200)^2$, shown by green circles. Here $log(FoR) = -5.96 + 2.0 logV + 2.35 logD + 0.39 logR + 0.61 logM$.

Graphically, Fig. 2 illustrates the fit between the temporal evolution for one of the models, Model B, and the corresponding values of the FoR model. It is important to note that the parameters of the FoR models may differ depending on the temporal evolution model assumed, as should be expected.

|  |  | Models of Temporal Dynamics | | | | |
|---|---|---|---|---|---|---|
|  |  | A | B | C | D | E |
| FoR model parameters |  |  |  |  |  |  |
|  | $\alpha_1$ | 2.00 | 2.00 | 2.00 | 2.00 | 2.00 |
|  | $\alpha_2$ | 1.81 | 2.35 | 2.70 | 2.20 | 1.64 |
|  | $\alpha_3$ | 0.50 | 0.61 | 0.61 | 0.56 | 0.33 |
|  | $\alpha_4$ | 0.02 | 0.39 | 0.92 | 0.37 | 0.65 |
| Goodness of fit |  |  |  |  |  |  |

|  | R² | 0.942 | 0.955 | 0.950 | 0.955 | 0.961 |
|---|---|---|---|---|---|---|
|  | BIC | -303 | -326 | -315 | -318 | -334 |
|  | MAPE | 0.133 | 0.112 | 0.179 | 0.114 | 0.153 |

*Table 2. Summarized results of model fitting.*

The summary of results, together with characterization of fit, are shown in Table 2:

- Parameter $\alpha_1$ was held constant at the value of 2.0 by appropriate rescaling. This serves to provide a domain-appropriate interpretation of the FoR, as discussed in the next section.
- Parameters $\alpha_2, \alpha_3,$ and $\alpha_4$ differ for alternative temporal dynamics models, as expected.
- BICs differ between the models for temporal dynamics, in part because the models had different numbers of parameters.
- MAPE is the average of MAPE values computed for the following three conditions:
    - We computed the MAPE for the years 1800–1900, while using the FoR parameters fitted only to the data that would be available to a hypothetical forecaster in the year 1800.
    - We computed the MAPE for the years 1800–2015, also using the FoR parameters fitted only to the data that would be available to a hypothetical forecaster in the year 1800.
    - We computed the MAPE for the years 1900–2015, while using the FoR parameters fitted only to the data that would be available to a hypothetical forecaster in the year 1900.

## 5. Discussion of Results

In this section, to make the discussion more domain-specific, we rename our four attributes in a more mnemonic fashion, specifically, we use *V* to refer to the muzzle velocity $x_{i1}$, *D* for the maximum effective range with $x_{i2}$, *M* for the mass of the projectile $x_{i3}$, and *R* for the maximum rate of fire $x_{i4}$.

It is important to note that the parameters of the FoR model (see rows labeled $\alpha_k$ in Table 2) derived in conjunction with five different models for the temporal dynamics of the FoR are mostly qualitatively comparable but quantitatively differ appreciably. This supports our contention that in fitting a FoR model to the empirical data, a range of models need to be explored. Different temporal evolution models can differ in the goodness of fit.

One notable outlier is the value of $\alpha_4$ obtained in conjunction with the temporal dynamics model A. This parameter describes the contribution of the rate of fire *R* into the overall FoR. Such a low number is implausible from the domain perspective. It implies that *R* offers almost no contribution to the FoR as compared to other attributes, and it is simply not consistent with the great attention given to increasing *R* throughout the history of small-arms technologies. It is also inconsistent with the other four temporal evolution models, which yield relatively high values of $\alpha_4$. We, therefore, are inclined to exclude the results corresponding to model A from further consideration. It is particularly remarkable because model A is the popular exponential model, the most common form of Moore's law. It is also the assumption underlying the original ANR approach. This reinforces the point that in fitting a FoR model to the empirical data, a range of temporal evolution models need to be explored.

Of the remaining four temporal evolution models, we assess the results associated with model B as the best overall, in terms of parsimony of the model, BIC, $R^2$, and MAPE. Therefore the remainder of the paper focuses on the FoR model parameters obtained in conjunction with model B.

With these parameters, the FoR model (see Eq. 1) becomes the following formula:

$$FoR = (1.1 * 10^{-6})V^{2.0}D^{2.35}M^{0.61}R^{0.39}, \tag{4}$$

where the velocity $V$ is in meters per second, effective range $D$ is in meters, bullet mass $M$ is in kilograms, and rate of fire $R$ is in number of rounds per minute. This is the FoR formula reflected in Fig. 2.

The parameters of the temporal evolution model B, optimized jointly with those of the FoR model, are given in the following formula (as shown in Fig. 2, where we elected to start the values of the FoR at 1.0 in the year 1200):

$$log(f_i^2) = 1.0 + 8.27 * 10^{-6}(t_i - 1200)^2 \tag{5}$$

To give Eq. 4 a more intuitive interpretation, let us rewrite it as follows:

$$FoR = (2.2 * 10^{-6})(0.5 * MV^{2.0})D^{2.35}M^{-0.39}R^{0.39}, \tag{6}$$

The second factor in the formula is simply the muzzle kinetic energy in joules. The remaining three factors can be considered as "corrections" for the effective range, mass of the projectiles, and rate of fire.

Specifically, the third factor in Eq. 6 can be interpreted as the correction for effective range. The exponent for this correction is 2.35, indicating a strong influence of the effective range on the FoR. Indeed, an increased engagement range is a trend consistent throughout the history of warfare, driven by the desire to reduce the casualties among one's own troops [Lawrence 2017].

The fourth factor of Eq. 6 can be interpreted as the correction for the mass of the projectile. The exponent is negative, indicating that greater mass (other than the one already included in the kinetic energy) is undesirable as it limits the amount of ammunition the infantryman can carry and use in the battle.

Finally, the fifth factor of Eq. 6 can be interpreted as the correction for the maximum rate of fire. The exponent is positive, indicating, as expected, that a higher rate of fire is a positive contributor to the FoR.

Overall, these results, obtained with ANR, fit well the empirical data over a great span of history and diversity of technologies, and are consistent with domain-specific expectations of these technologies. Thus, we show empirically that ANR can be used to analyze the evolution of technology over an extremely long period of time, when applied to functionally similar but otherwise widely different technology families, and in a case of superexponential growth of technology characteristics. To our knowledge, the use of ANR has not been explored before for such conditions.

A disclaimer worth making here is that Eq. 6 should not be mistaken for a design guide or a comprehensive model of a small-arms performance. One cannot arrive at a better weapon merely by increasing the weapon's FoR. Much more than just the FoR goes into the good design of a weapon.

## 6. Using the Models for Analysis and Forecasts

In this section, we do not aim to recommend or advocate any technology development directions. Rather we wish to illustrate how a technology analyst could use the models we derived in the preceding sections.

First, the analyst should determine the quantitative extent of improvements in the FoR that are likely to occur in infantry small arms by the year 2050. Second, they should explore whether the anticipated increment of the FoR is feasible in terms of current research and development directions. Note that the following discussion is based strictly on information available in the open literature.

To answer the first question – how much will small arms improve by the year 2050 – we use the temporal dynamics model, Eq. 5, which gives us the value of *log*FoR=6.97 in the year $t_m = 2050$. (It is convenient here to use the logarithm of the FoR.) The highest *log*FoR found in our data for currently known weapons (see Fig. 2) is 6.18. This means that *log*FoR is likely to grow by about 0.80.

The next question is whether we have any evidence that such a substantial increase in *log*FoR is feasible. If not, then our forecast might be doubtful. One way to answer the question is to see whether the currently known research and development (R&D) directions might lead to such an increment of the FoR by the year 2050, assuming no major discontinuity in the technology of small arms. Specifically, are the currently known directions of R&D in small arms likely to raise *log*FoR by 0.80?

To answer this question, consider the FoR model, Eq. 5, and begin with the term that refers to muzzle velocity. There are expectations [Magnuson 2018] that the Army near-future rifle will exhibit a higher muzzle velocity than current rifles. In the last 100 years, the muzzle velocity of various rifles ranged from about 800 to 1000 m/s. Assuming a plausible increase of 10% and using Eq. 5, this would yield an increase of 2.0·*log*(1.1) = 0.08 in the value of *log*FoR.

The next variable term of Eq. 5 refers to the effective range of a weapon. Here we note developments toward computerization (with elements of artificial intelligence) of aiming and fire control of an infantryman weapon; see, e.g., National Interest [2019]. Commercially available examples of such technologies already exist [Boyd and Lupher 2013]. These seem to imply that the effective range (in terms of the probability of hit and assuming the projectile [e.g., a bullet] retains sufficient energy at the target) could be increased by 50–100%, as compared to today's small arms. If we were to assume a 70% increase in D, *log*FoR would increase by an increment of 2.35·*log*(1.7) = 0.54.

The third variable term of Eq. 5 refers to the projectile mass. Recent years have seen a move toward a heavier bullet, such as the 6.8mm round [National Interest 2019], with the bullet mass perhaps 80% greater than in the current NATO-standard 5.56 × 45mm cartridge. With such an increase in bullet mass, *log*FoR would increase by an increment of 0.61·*log*(1.8) = 0.16.

Finally, the fourth variable term of Eq. 5 has to do with the cyclic rate of fire. In the current literature, we do not see significant indications of developments toward a greater rate or arguments that such an increase might be desired. Thus, we cannot forecast that an increase in this rate is a likely contributor toward the growth of the FoR by the year 2050.

Having considered in turn all four variable terms of Eq. 5, we can summarize that the currently known developments in small-arms technology might yield a total potential increment of *log*FoR about 0.08 + 0.54 + 0.16 = 0.78. This is very close to the total increment of 0.80 we forecast based on the long-term trend. It does *not* mean we recommend or advocate these developmental directions, or forecast that they will be successful. Nevertheless, the fact that the developments discussed in the open literature appear, in principle, capable of achieving the forecasted increment strengthens the forecast feasibility.

# 7. Conclusions and Recommendations

We empirically explore an approach – a generalization of ANR – to deriving a technology's FoR from the data of the technology's history. A FoR follows a regular temporal trend and as such is useful for technological analysis and strategic forecasting, particularly for those technologies for which a suitable performance measure cannot be formulated.

We further demonstrate, in our case study, that an *a priori* assumption of a single exponential law can be misleading. Instead, multiple temporal evolution models should be considered, and these should be fitted to the data jointly with the FoR model.

We also show empirically, using a demanding case study, how the models derived via the proposed approach can be used to (a) produce a quantitative long-term forecast of increases in a technology's FoR, and (b) assess quantitatively the extent to which a portfolio of R&D efforts supports the forecasted increase. The ANR approach is conceptually and computationally parsimonious, with results that are visually clear and explainable – an important aspect in communicating with technology management.

Although in this paper the study of long-term trends in small arms serves mainly as an illustrative case, our findings for this technology are also valuable. The FoR of small arms derived in this paper exhibits a very long-term – since at least 1200 CE – superexponential trend, a kind of Moore's law. To our knowledge, it is the first time that such a trend or law has been reported for small arms. This is also the longest, to our knowledge, exponential or superexponential trend reported for any technology. The existence of such long-term regularities is in itself remarkable and relevant to the study of technology analysis.

The lessons of our research lead us to offer several recommendations to technology analysts and R&D managers. One insight is that longer-term trend analysis may be beneficial, as it is likely to reveal the appropriate, more generally applicable attributes of the relevant technology families. It may also help avoid a bias toward particular design features or development directions that happen to dominate a shorter-term history.

On a related note, analysts should consider exploring a broader range of functionally similar families of systems, even if the underlying mechanisms are very different. For example, the relevance of crossbows or smoothbore muskets may not be obvious to a study of trends for modern assault rifles. Nevertheless, broadening the historical period and the set of underlying technologies may yield insights that are more robust.

Using approaches such as the one discussed in this paper, technology analysts should attempt to validate their forecasts with a quantitative assessment of whether a portfolio of research efforts – actual or proposed – is likely to reach the forecasted improvements in a FoR or key performance attributes of the technology in question.

## Acknowledgments

The authors thank James Newill for insights on modern rifles, Louise McGovern for assisting with finding the sources of data, and Carol Johnson for editing the manuscript.